# Cloud boundary height measurements using lidar and radar

Victor Venema[1], Herman Russchenberg[1], Arnoud Apituley[2],
Andre van Lammeren[3], Leo Ligthart[1]

[1] IRCTR, Delft University of Technology, Delft, The Netherlands
[2] National Institute of Public Health and the Environment, RIVM
[3] Royal Netherlands Meteorological Organisation, KNMI

**Abstract.** Using only lidar or radar an accurate cloud boundary height estimate is often not possible. The combination of lidar and radar can give a reliable cloud boundary estimate in a much broader range of cases. However, also this combination with standard methods still can not measure the cloud boundaries in all cases. This will be illustrated with data from the Clouds and Radiation measurement campaigns, CLARA. Rain is a problem: the radar has problems to measure the small cloud droplets in the presence of raindrops. Similarly, few large particles below cloud base can obscure the cloud base in radar measurements. And the radar reflectivity can be very low at the cloud base of water clouds or in large regions of ice clouds, due to small particles. Multiple cloud layers and clouds with specular reflections can pose problems for lidar. More advanced measurement techniques are suggested to solve these problems. An angle scanning lidar can, for example, detect specular reflections, while using information from the radars Doppler velocity spectrum may help to detect clouds during rain.

## 1    Introduction

The main objective of the three Dutch Clouds and Radiation (CLARA) campaigns in 1996 was to increase the understanding of radiative processes in the atmosphere by making high quality cloud measurements (Van Lammeren, 1998). The instrumentation in Delft (close to the Dutch coast) included: lidars, radar, infrared radiometer, microwave radiometer, and radiosondes. During extended fields of water clouds an aircraft performed in situ measurements of the drop size distributions with an FSSP-100.

For cloud boundary measurements a main advancement is the synergetic use of lidar and radar, a combination that is also planned to be used in e.g. the European Earth Radiation Mission. For water clouds the radar is normally best at measuring the cloud top and the lidar at measuring the base. However, there are still situations when cloud boundaries are difficult to measure. The problem is often the lack of detectable cloud reflections in one of the instruments, so that the synergy can not be used. When the lidar signal is totally attenuated, for example, the radar will have to measure both the boundaries of possible higher clouds alone. Based on some case studies of low and mid altitude clouds this article will argue that *current* radar measurements are not always up to this task, due to problems with very small or very large particles.

During rain, radar cannot measure the radiatively most important (cloud) particles, with current measurement techniques, as the precipitating particles dominate the signal. Then lidar would have to measure both boundaries alone, which is often not possible. Ice clouds giving specular reflections might lead to erroneous interpretations of the cloud boundaries for lidar. For all the above measurement problems new measurement techniques have to be developed and at least the difficulties should be recognised.

Ignoring the problems can lead to large errors and biases. For example, in the Netherlands it rains more than 0.1 mm/hr about 7 percent of the time (KNMI, 1992). A measurement technique that does not recognise rain can make large errors and just ignoring the rain cases may introduce a bias.

This article aims at improving the understanding of the radar-lidar measurements of cloud boundaries. At least a qualitative understanding of the micro-physical cloud properties and scattering is necessary to understand the cloud boundary measurements. This article will focus on some situations that can be difficult to measure, illustrate them with measurements and suggest a direction for new measurement techniques.

## 2 Instruments

### 2.1 Radar

The Delft Atmospheric Research Radar (DARR) is a 9-cm Frequency Modulated Continuous Wave (FM-CW) Doppler radar. The radar measurements are averaged over 5 s, the beam width is 1.8°, the sensitivity at half the maximum range is about -27 dBZ and the range resolution can be set to 15 or 30 m, giving a maximum range of respectively 4 or 8 km. For small randomly distributed water droplets the received power is proportional to the diameter to the sixth power. As the wavelength of DARR is much longer than that of typical cloud radars, some of the reflections may be due to spatial refractive index variations, caused by turbulence. In the CLARA database we estimate that 13 % of the time the coherent air scatter (clear-air scatter) is more than -20 dBZ in the boundary layer. This means that it may be stronger than reflections from clouds, thus fair weather cumulus may be masked for cm-wave radar. Also variations in mass density of particles on spatial scale of half the wavelength – coherent particle scatter – can enhance the reflection strength of these particles for cm-wave radars (Erkelens *et al.*, 1999 and Venema et al., 1999), this could especially be important at edges of clouds.

### 2.2 Lidars

Two different near-infrared backscatter lidar systems were used: The Vaisala CT-75k lidar ceilometer (wavelength 906 nm) and an experimental system (wavelength 1064 nm was used): the high temporal resolution lidar (HTRL lidar). The Vaisala is a commercial system, having a range resolution of 30 m and an integration time of 12 s, the pulse repetition rate is 5.1 kHz with an energy of 1.6 mJ. The HTRL lidar stores the single-shot returns at a rate of 1.6 Hz, the pulses of 10 ns have an energy of 0.3 J, and the lidar has a range resolution of either 1.5 or 7.5 m. More information on the HTRL lidar can be found in Apituley (1999). It would be preferable to use lidar to measure cloud boundaries, as it uses light. However, attenuation of the laser beam is very important for lidar. So for a cloud with a high optical depth (typically above about 3), radar measurements are needed, for instance to measure the cloud top height.

The range output of the Vaisala and DARR was intercalibrated on a far away chimney and was correct within one range cell. A comparison between the measurements of the lidar systems shows that the range of the HTRL lidar is equally accurate. The instruments were placed within 15 m of each other. The measurement times were synchronised afterwards. In all case studies presented in this paper the instruments were pointed to the zenith.

## 3 Observed phenomena

Ideally, the cloud boundary should be derived from the micro-physical cloud properties (extinction coefficient or liquid water content) to be most useful for climate studies. At this stage, however, it is not feasible to routinely convert remote sensing measurements of clouds into micro-physical cloud properties.

The approach taken in this article is to use directly the measured reflection profiles. The height at which the signal decreases considerably will be called cloud top or base. This qualitative approach is sufficient for this article, as the kind of problems treated are not solved by using a more refined algorithm on the same reflection profiles. Note that with this definition to cloud top and base do not have to correspond to the true (radiatively significant) cloud boundaries.

In this section five cases studies with radar-lidar cloud measurements will be presented and qualitatively discussed in terms of the micro-physical cloud properties which are relevant for the retrieval of cloud boundaries. Based on the interpretation of these cases some suggestions for new measurement techniques will be given in section 4.

3.1 Effect of particle size

In general, one can say that radar reflection measurements are dominated by the large particles and the lidar measurements by the small particles. The importance of this depends on the width of the particle size distribution.
The radar receives reflections from below the cloud base that is measured by lidar in the ice cloud measurement shown in fig. 1. The cloud base of the lidar is in this case likely to be representative. The cloud base measured by radar is thus probably 100 to 500 m too low. A likely cause of these radar reflections are some sparse falling crystals (the radar velocity is between 0.5 and 1 m/s downward at the radar cloud base). Weitkamp *et al*. (1999) also observed with their 3-mm wavelength cloud radar a base which was 100 to 600 m too low for an altocumulus. The base measured by DARR of clouds containing ice is often observed to be lower than the base measured by lidar. It seems to be a typical error for ice clouds that the radar base is a few hundred metres too low. For DARR this may also be explained by coherent air scatter (Bragg), see Rogers and Brown (1998), but for this case that seem unlikely as coherent air scatter should have a Doppler velocity around 0 m/s. Note that the cloud (fig. 1) is called an ice cloud because the radar reflection is dominated by ice crystals. However, it can also be a mixed cloud, and the lidar may receive backscatter from water droplets. In the entire cloud the structures seen by radar are more vertical – indicating large falling particles – and by lidar more horizontal – indicating small floating particles.

Another example of the effect of large particles on radar measurements is shown in fig. 2. This measurement of virga (precipitation that does not reach the ground) was made on November 27, 1996. The lidar backscatter (fig. 2a) shows a thin stratus layer at 2.5 km, with ice crystals precipitating out of this cloud and evaporating above 1.5 km. The radar (fig. 2b) only sees the reflections of the large falling ice crystals; the presence of the cloud at 2.5 km does not even increase the radar reflections significantly. The radar 'cloud base' is in this case placed almost one kilometre too low. A distinction between the small cloud particles and large falling crystals is possible by using information present in the Doppler spectrum; a good example is shown in fig. 2c. The cloud seen by lidar at 2.5 km is revealed by the small radar reflections with a positive upward velocity.

The contrast in particle size is also very large in rain. Figure 3 shows a very light rain event (December 6, 1996), which at its peak is no more than 1 mm/hrs and the drops have a maximum fall speed of 5 to 6 m/s. The three stratus clouds present in the lidar measurement (fig. 3a) are not distinguishable in the radar measurement (fig. 3b). In the measured velocity it can be seen that the reflections from the precipitation dominate the total reflection so strongly that the *average* velocity also does not reveal the position of the clouds. A retrieval of the cloud boundaries by current radar techniques is thus not possible.

Ice clouds can also contain very small ice crystals, especially at the cloud top. In the second part of the ice cloud radar measurement shown in fig. 4a, the radar does not detect most of the cloud. This is probably due to the particles being too small here, as the cloud is gradually decaying. The radar would make an error in both the cloud top and base in the last part of this measurement. Weitkamp *et al.* report about a total multi-layered cirrus cloud which their cloud radar could not detect as the reflectivity was below -30 dBZ.

One CLARA measurement of stratocumulus is shown in fig. 5a; for this cloud in situ FSSP data is available as well, see fig. 5b. The cloud base as measured by lidar is about a 100 meters below the first radar signals from DARR. This is because the radar reflection from the small droplets (just 10 μm) at cloud base is much too small, as was confirmed by calculating the radar reflectivity belonging to the drop size distributions from fig. 5b. Based on a simple model for a stratus water cloud Sassen *et al.* (1999) estimate that a radar with a minimum detectable signal of -30 dBZ will detect the first signals 200 m above the lidar cloud base.

3.2     Effect of attenuation

Although for 95 Ghz radar attenuation can be a problem (Danne *et al.*, 1999), for cm-wave radars the attenuation can be neglected. In the case of lidar, it is common that the clouds (especially clouds containing water) are optically too thick for the light to penetrate to the cloud top. Another problem is that from the measured lidar backscatter profiles it is hard to determine whether the signal decreases towards the noise level due to total attenuation or because the top of the cloud is reached.

In the ice cloud shown in fig. 1, the cloud 'top' measured by lidar is often a few hundreds of meters lower than the radar cloud top, especially between 21 and 22 hours. This height difference is likely to be caused by attenuation of the lidar signal.

In the lidar measurement of light rain (fig. 3a) three cloud layers are present. The upper two layers are not visible in the last part of the lidar measurement due to attenuation by the lowest cloud, while at least the cloud producing the precipitation should still be present. As the radar can not measure the clouds, due to the rain, there is no information on the cloud base of the two upper clouds and no information on the cloud top of the lower two clouds.

The stratocumulus cloud measured on the 19th of April 1996 shows a peak at different heights for lidar and radar (fig. 5a). During most of the measurement the lidar receives no power from the region where the radar sees the cloud. The bumpy character of the cloud makes it hard to analyse. One can only speculate whether the lidar power was attenuated in the region of the radar reflections or whether the radar is receiving reflections from above the cloud (maybe clear air scatter).

3.3 Effect of specular reflections

Non-spherical ice crystals can reflect light almost like a mirror; this can cause specular reflections in the vertical direction due to horizontally aligned ice crystals. The largest dimension of the crystal can be horizontally alignment due to aerodynamical forces. Thomas *et al.* (1990), for example, measured an angular dependence in lidar echoes of a cirrus cloud of just 0.3° around the zenith. Likewise, theoretically a decrease in the vertically backscattered power should result when the ice crystals are no longer aligned horizontally, compared to the aligned case.

A lidar measurement of an ice cloud on the 25th of March 1998 (fig. 4) shows a dark band at about 4.3 km height. The cloud and its dark band (a layer of reduced backscatter) lasted about 4 hours. Only the last hour is shown, as for this part radar data was available as well. The first two hours the dark band was 200 to 300 m wide, the last two hours about 100 m. Fall streaks with high lidar and radar echoes fall from the top part, through the dark band into the lower part. The width of the velocity spectrum is about 3 times larger in the dark band than in its environs; an indication of turbulence.

Data from a radiosonde at 6 hrs UT revealed a strong change in wind direction. At 4 and 4.6 km the wind direction was North, but in the dark band it was East; In the dark band the wind direction was constant over 200 m, the wind speed 1 m/s and the temperature was -17 °C. The radiosonde recorded a relative humidity from 75% at 4 km to 90% at 4.5 km. The presence of liquid water is thus not likely.

A possible explanation for this measurement is that in the upper and lower part the lidar backscatter are high due to specular reflections from horizontally aligned ice crystals and that in the dark band these crystals are no longer horizontally aligned due to turbulence. Unfortunately, there was no lidar measurements with a beam that is directed a few degrees away from the zenith. This would be needed to ascertain whether the lidar backscatter was specular or not.

Another possibility is that it are simply two separate clouds, with a cloud-free region in between. The fall streaks could then connect the clouds by cloud seeding: crystals falling from the top cloud into the lower cloud. However, if this were true the lower cloud should contain water (because of the cloud seeding), which is unlikely, given the radiosonde data. Concluding, specular reflections seem most probable explanation, but it should be directly measured to be sure.

## 4  Proposed advanced measurement techniques

In a large number of cloudy conditions the current measurement techniques suffice to measure cloud geometry. For example, in the first case study of an ice cloud (fig. 1) the cloud boundaries are probably representative for the true boundaries, taking the lidar cloud base and the radar cloud top. However, to achieve representative operational measurements under more cloud conditions, this sections gives some ideas for improved measurement techniques which would have been useful in the previous case studies, with no attempt of being comprehensive.

### 4.1  Lidar

The measurement artefacts created by specular reflections should disappear when the lidar is tilted under a small angle. Tilting the lidar will also enhance the contrast between cloud droplets and raindrops, due to the higher backscatter of raindrops in the vertical direction (Venema *et al.*, 1998). For the measurement of the micro-physical properties and process studies of clouds specular reflections can be interesting. In these cases a lidar that can automatically scan in angle would be useful; in the case of the dark band (fig. 4) it could ascertain the specular character of the backscatter. Experiments with a scanning lidar would be needed to see if specular reflections are a significant problem for the measurement of cloud geometry.

When a lidar receives molecular (Rayleigh) backscatter from a region above the cloud, this can be used to estimate whether the lidar was totally attenuated or not and thus whether the cloud top was measured reliably by lidar. Had Rayleigh scatter been measured above the stratocumulus cloud in fig. 5, this would have simplified the interpretation.

### 4.2  Radar

Using radar for cloud geometry measurements during rain (including rain that does not reach the ground) can be made possible by using information of the Doppler velocity spectra. In the case shown in fig. 2c it would suffice to store the reflection of the particles going up. In the measurement of light rain (fig. 3) the cloud may be made visible by looking at the power of the upward moving particles, even though the *average* velocity has not been changed by the presence of the cloud. But as the velocity of the cloud particles is easily perturbed by updrafts and downdrafts, maybe Doppler polarimetry or Doppler multi-frequency spectra are needed to unambiguously identify the smallest particles. The radar should, furthermore, have a very high sensitivity, Sassen *et al.* (1999) estimate based on a simple cloud model that a radar with a sensitivity of -40 dBZ will give about the same cloud base height as a lidar for water clouds. For ice cloud the sensitivity sometimes needs to be even better (Sassen and Khvorostynov, 1998).

4.3 Sensor synergy

It would be useful to have a Doppler lidar next to a Doppler radar for some measurement campaigns. The fall velocity of the particles is a function of the size, next to the shape of the particles. The difference in fallspeed between lidar and radar may serve as an indirect measure of the width of the particle size distribution. If both lidar and radar measure about the same particle speed, one would have more confidence in interpreting the combined measurements in terms of one effective particle shape and size. In case of a difference in fall speed, the fall velocity of the lidar (small particles) can be used by a algorithm that detects cloud boundaries in the radar Doppler spectrum.

## 5   Concluding remarks

To make accurate and useful cloud boundary measurements in a broad range of atmospheric conditions, one has to combine lidar and radar. This article presented a number of case studies to illustrate the physical processes important for measurements of cloud geometry: the influence of particle size, attenuation and specular reflections. The conclusion drawn from analysing these case studies is that the measurement techniques still have to be improved a great deal before representative operational cloud measurements under all atmospheric conditions are feasible. Difficult conditions are: rain and virga, ice clouds with a broad size distribution and ice clouds with specular reflections. How often these problems occur, should be the subject of further studies and optimised measurement techniques have to be developed for these cases.

**Acknowledgements**
The authors would like to thank Christine Unal for calculating the Doppler spectra of virga, and Gerard Kos for making the FSSP measurements. This work was partly funded by the Dutch National Research Programme on Global Air Pollution and Climate Change (NOP).

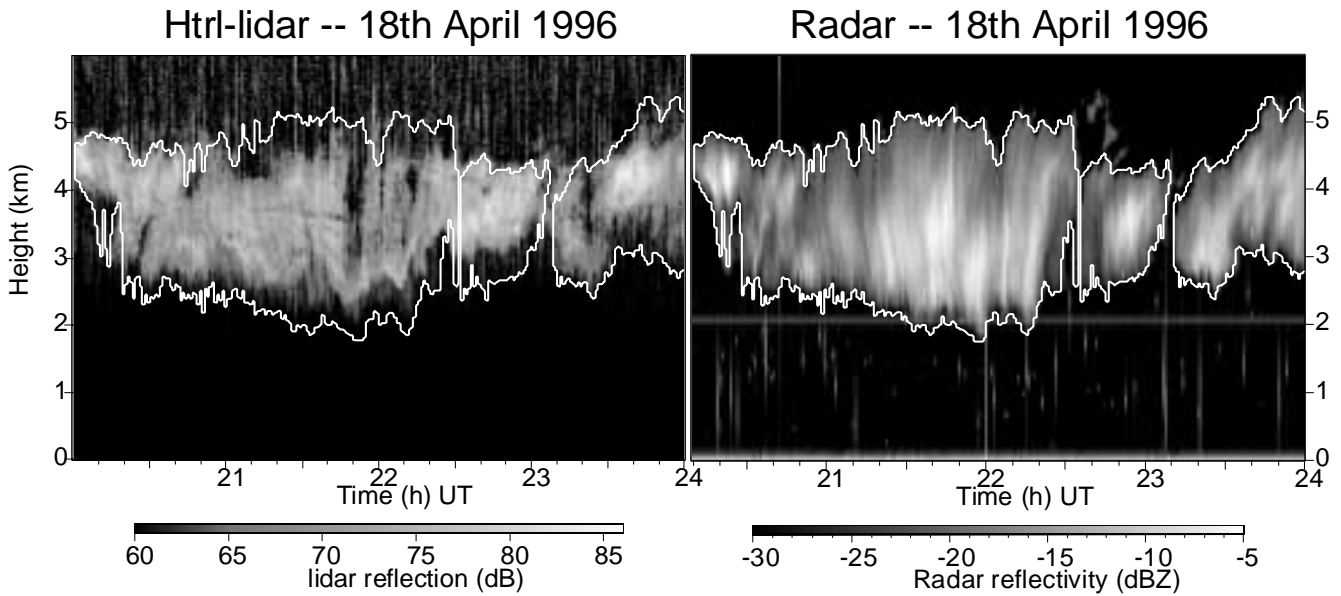

**Fig. 1.** Measurement of the vertical reflection profiles of an ice cloud on the 18th of April 1996. Fig. 1a is the HTRL lidar range corrected backscatter in arbitrary dB units. Fig. 1b gives the equivalent radar reflectivity factor measured by DARR. To facilitate comparison, the contour of the cloud as measured by radar is plotted in both figures. The dots and vertical lines in fig. 1b are point targets of unknown origin, also called angels. The horizontal line at 2.1 km in fig. 1b are an effect of the radar system, and should be ignored.

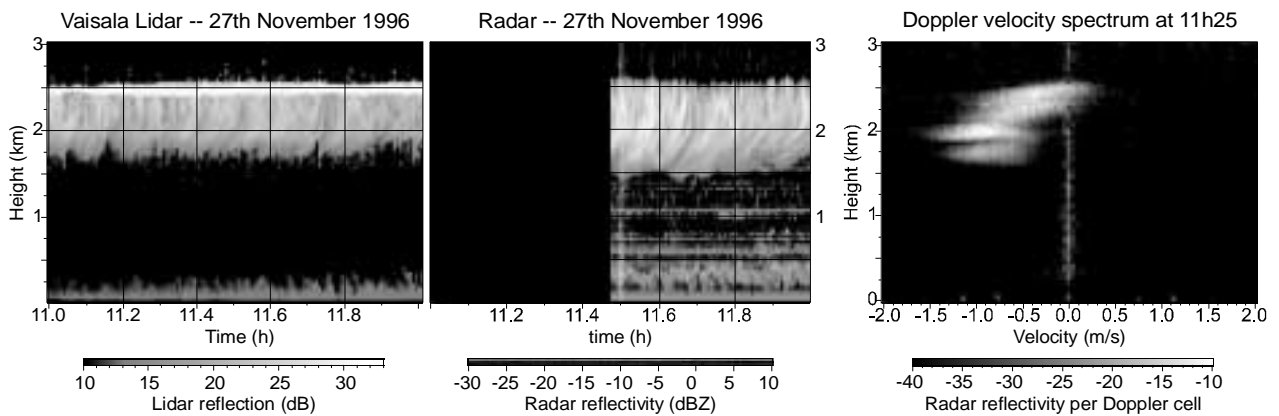

**Fig. 2.** A measurement of virga on the 27th of November 1996. Fig. 2a was measured by the Vaisala lidar in range-corrected dB units. It shows a stratus layer at 2.5 km and ice crystals precipitating out of this layer, which evaporate above 1.5 km. Fig. 2b shows the same period for the 9-cm radar, which measures the falling crystals between 1.5 and 2.5 km. In the first half of this measurements the radar was in another mode to measure some Doppler velocity spectra. One of these in presented in fig. 2c: the radar reflectivity velocity spectra are plotted as a function of height. The unit is the effective radar reflectivity factor per Doppler cell (of 4.7 cm/s) in dBs. The vertical line of reflections at zero velocity is ground clutter.

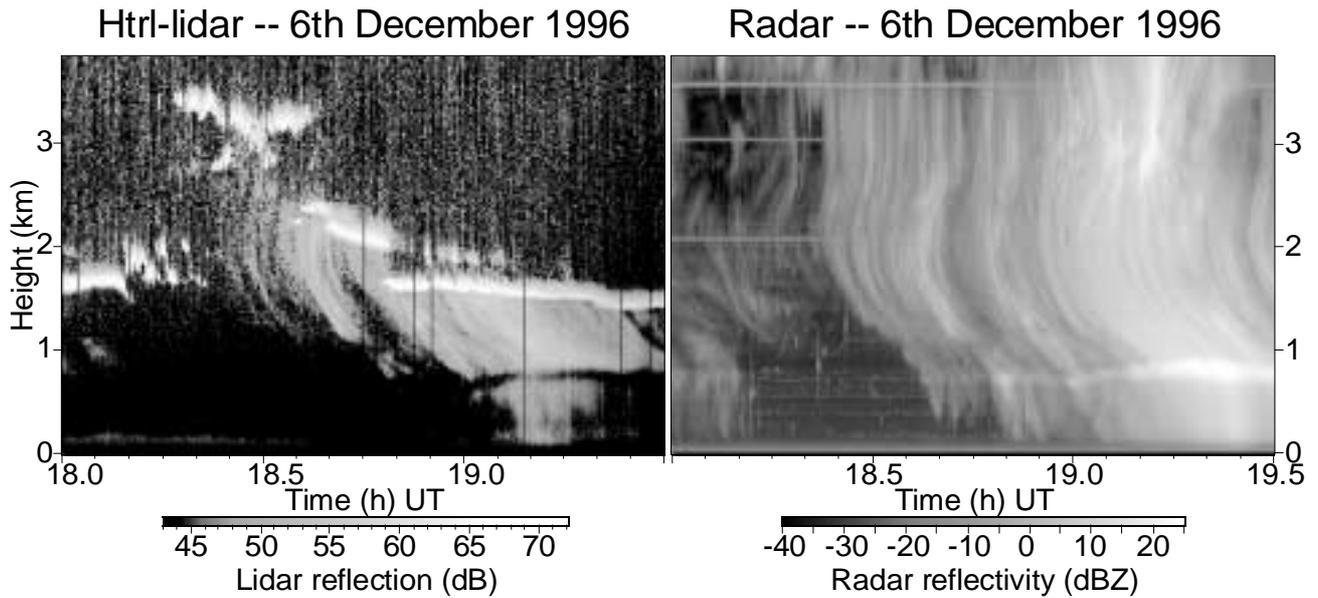

**Fig. 3.** This measurement of light rain (on the 6th of December 1996) shows a large difference between the HTRL lidar (fig. 3a) and the radar (fig. 3b). The multiple stratus layers visible in the lidar measurement are not present in the radar, which only sees the reflections from the falling crystals. The 0-degree isotherm is at 800 m, which is revealed by the traditional bright band in the radar and a dark band in the lidar. This dark band is discussed in Venema *et al.* (1998) and Sassen and Chen (1995).

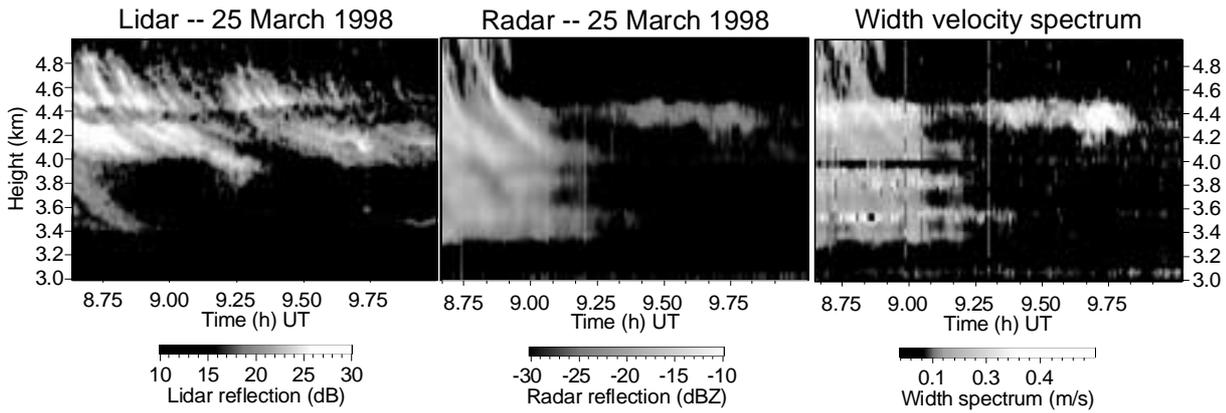

**Fig. 4.** A thin dark band is visible in the middle (at 4.3 km) of the lidar measurement of an ice cloud which lasted for 4 hours on the 25th of March 1998, only the last hour is shown in fig. 4a. The dark band is about 10 dB deep. Fig. 4b is the radar reflectivity; after 9.25 hrs the particles are probably too small to be measured by radar. Fig. 4c shows the Doppler width of the radar signal; this is the standard deviation of the velocity spectrum (such a spectrum as a function of height is shown in fig. 2c). This Doppler width is about three times as large at the position of the dark band. In the dark band itself there are still some reflections after 9.25 hrs (see fig. 4b), which could either be some residual particles, or coherent scattering due to the turbulence. The very straight horizontal lines in the Doppler width at 3.95 and 3.5 km are a system effect.

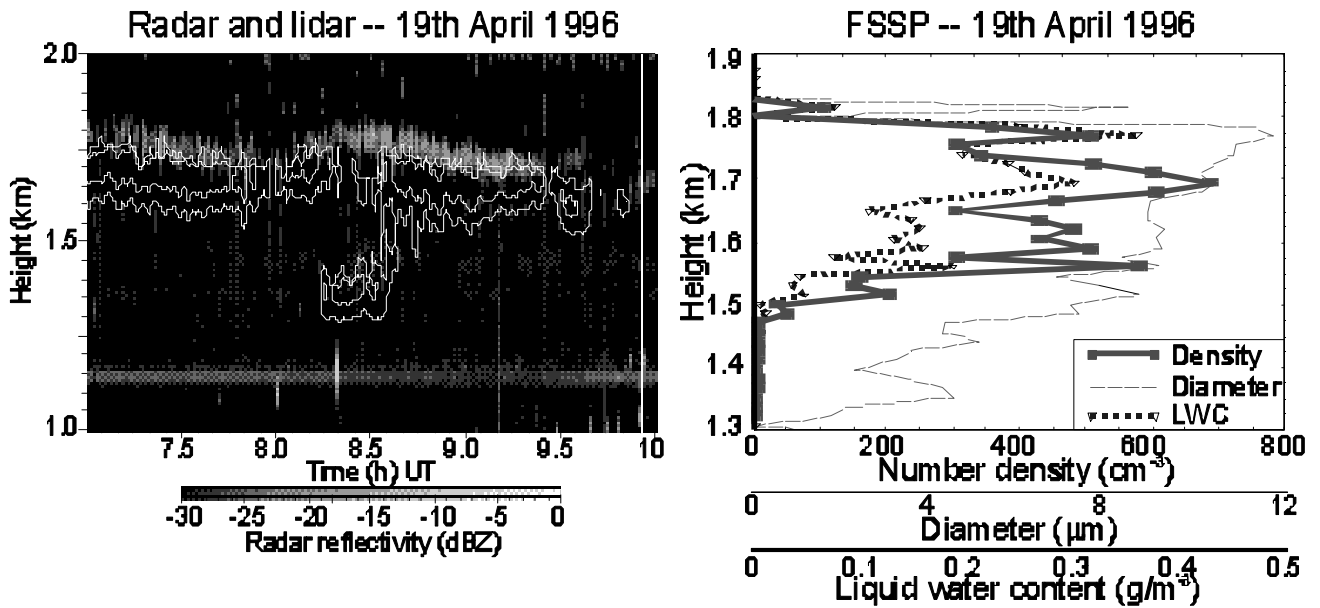

**Fig. 5.** A stratocumulus cloud measured with radar and lidar (Vaisala) is shown in fig. 5a. The background of the figure is the radar reflectivity and the contours the range corrected lidar backscatter. The contours are 10 dB apart. The peak of the vertical profile of the radar is significantly above the peak of the lidar. The radar receives power from a region where the lidar does not receive any backscatter. Figure 5b shows the averaged number of droplets, the average drop size and Liquid Water Content as measured with an FSSP-100. For these profiles all data from a 3-hour measurement in a region 50 km around Delft was used.